\documentclass{amsart}
\usepackage{amssymb}
\usepackage{amsmath}
\usepackage{amsfonts}
\usepackage{graphicx}
\usepackage{epstopdf}

\newcommand{\beq}{\begin{equation}}
\newcommand{\eeq}{\end{equation}}

\begin{document}

\begin{center}
{\Large \bf On a Gopakumar-Vafa form of partition function of Chern-Simons theory on classical and exceptional lines} \\
\vspace*{1 cm}
{\large  R.L.Mkrtchyan 
}
\vspace*{0.5 cm}

{\small\it Yerevan Physics Institute, 2 Alikhanian Br. Str., 0036 Yerevan, Armenia}

\end{center}\vspace{2cm}

{\small  {\bf Abstract.} We show that partition function of Chern-Simons theory on three-sphere with classical and exceptional groups (actually on the whole corresponding lines in Vogel's plane) can be represented as ratio of respectively triple and double sine functions (last function is essentially a modular quantum dilogarithm). The product representation of sine functions gives  Gopakumar-Vafa structure form of partition function, which in turn gives a corresponding integer invariants of manifold after geometrical transition. In this way we suggest to extend gauge/string duality to exceptional  groups, although one still have to resolve few problems. In both classical and exceptional cases an additional terms, non-perturbative w.r.t. the string coupling constant, appear. The full universal partition function of Chern-Simons theory on three-sphere is shown to be the  ratio of quadruple sine functions. We also briefly discuss  the matrix model for exceptional line.}


\section{Introduction}

In the present paper we continue investigation of integral representation of universal partition function of Chern-Simons theory on 3d sphere with an arbitrary gauge group, derived in \cite{M13}. Partition function is universal in the sense that it depends on gauge group through parameters on Vogel's plane (two-dimensional projective plane, factorized over all permutations of its projective parameters $\alpha, \beta, \gamma$), at particular values of which given in Table \ref{tab:1} one get an answers for the corresponding simple Lie algebras \cite{V0}. So, this representation not only gives partition functions of Chern-Simons theory on 3d sphere with any simple Lie gauge group, but also extend them to the whole Vogel's plane of universal parameters. Such an extension of $SU(N)$ theory to arbitrary $N$ is necessary for  $1/N$ expansion \cite{H1} and then for gauge/string duality. In present paper we shall use that extension to construct an analog of $1/N$ for exceptional groups.

\begin{table}[h] \caption{Vogel's  parameters for simple Lie algebras}     
\begin{tabular}{|r|r|r|r|r|r|} 
\hline Algebra/Parameters & $\alpha$ &$\beta$  &$\gamma$  & $t=\alpha+\beta+\gamma$ & Line \\ 
\hline $SU(N)$ & -2 & 2 & $N$ & $N$ & $\alpha+\beta=0 $\\ 
\hline $SO(N), Sp(-N)$ & -2  & 4 & $N-4$ & $N-2$ & $\alpha+2\beta=0$ \\ 
\hline $Exc(n)$ & -2 & $n+4$  & $2n+4$ & $3n+6$& $\alpha+\beta+\gamma=0$ \\ 
\hline 
\end{tabular} \label{tab:1}
\end{table}

In Table \ref{tab:1} for $SU(N)$ and $SO(N)$ $N$ is positive integer, for $Sp(-N)$ $N$ is negative even integer, for exceptional line $Exc(n)$ $n=-1,-2/3,0,1,2,4,8$ for  $A_2,G_2, D_4, F_4, E_6, E_7, E_8$  respectively. Note also that $N \rightarrow -N$ transformation for classical groups corresponds to transposition $\alpha \leftrightarrow \beta$.

Consider partition function of Chern-Simons theory on 3d sphere:
\begin{eqnarray}
Z=\int DA \, exp\left(\frac{i\kappa}{4\pi}\int_{S^3} Tr \left( A \wedge d A + \frac{2}{3} A\wedge A \wedge A \right)\right)
\end{eqnarray}
Here $Tr$ means an arbitrarily normalized invariant bilinear form in simple Lie algebra of compact gauge group. So, rescaling of bilinear form leads to rescaling of $\kappa$ (and $\alpha,\beta,\gamma$).

This quantity is calculated by Witten in  \cite{W1} and answer, with a natural  choice of framing, appears to be the $S_{00}$ element of modular transformations matrix $S$. It is transformed in \cite{M13} into:

\begin{eqnarray}\label{totalfree}
-\ln Z =
(dim/2)\ln(\delta/t)+
\int^{\infty}_0 \frac{dx}{x} \frac{f(x/\delta)-f(x/t)}{(e^{x}-1)} \label{Ftotal}
 \end{eqnarray}

with

\begin{eqnarray}\label{gene}
f(x)&=&\frac{\sinh(x\frac{\alpha-2t}{4})}{\sinh(x\frac{\alpha}{4})}\frac{\sinh(x\frac{\beta-2t}{4})}{\sinh(x\frac{\beta}{4})}\frac{\sinh(x\frac{\gamma-2t}{4})}{\sinh(x\frac{\gamma}{4})} \\ \label{dim}
dim &=& \frac{(\alpha-2t)(\beta-2t)(\gamma-2t)}{\alpha\beta\gamma}\\ 
\delta&=&\kappa+t \\
 t&=&\alpha+\beta+\gamma
\end{eqnarray}

Here Vogel's parameters  $\alpha, \beta, \gamma$ together with $\kappa$ combine into a set of four projective parameters, characterizing simple Lie algebra and level $k$ of corresponding affine untwisted Kac-Moody algebra, $\kappa$ serving as a coupling of Chern-Simons theory.  Their correspondence with simple Lie algebras is given in Table \ref{tab:1} in the so called minimal normalization characterized by the square of long root being equal to 2. Then $t$ becomes  dual Coxeter number and  $\kappa$ becomes integer level $k$. Function $f(x)$ ("universal character of adjoint representation"), derived in this universal form in \cite{MV1}, is the character of adjoint representation on the line $x\rho$, where $\rho$ is the Weyl vector in roots space, $x$ an arbitrary parameter. $\delta=\kappa +t$ is (in an arbitrary normalization) usual effective coupling constant equal to the sum of bar coupling $\kappa$ and half of eigenvalue $2t$ \cite{V0} of second Casimir operator in adjoint representation.

Besides partition function, there are some other universal quantities in Chern-Simons theory, see \cite{MV1}, e.g. central charge

\begin{eqnarray}\label{cg1}
c=k\frac{ dim}{k+h^\vee}=\frac{\kappa(\alpha-2t)(\beta-2t)(\gamma-2t)}{\alpha\beta\gamma(\kappa+\alpha+\beta+ \gamma)}=\\
\frac{(\delta-t)(\alpha-2t)(\beta-2t)(\gamma-2t)}{\alpha\beta\gamma\delta}.
\end{eqnarray}
unknot Wilson loop in adjoint representation, etc.

In paper \cite{M13} we noticed that if it was possible to extend an  integration range in the integral representation (\ref{Ftotal}), specialized to $SU(N)$ group,  to entire real axis, one can immediately obtain a Gopakumar-Vafa (GV) form \cite{GV} of corresponding dual topological string  plus additional terms non-perturbative w.r.t. the string coupling constant. In present paper we present the correct way for such an extension. Particularly, starting from representation of $SU(N)$ partition function in terms of triple and double Barnes' gamma functions, derived in \cite{M13-2}, we show that it can be expressed in terms of triple sine functions (see  \cite{Nar,Kur,KurKoy,Shi} for definitions of multiple sine and related functions), by simple formula: 

\begin{eqnarray} \label{SUN}
Z=\frac{\sqrt{d}}{ \sqrt{N}}  \frac{S_3(2N+2|2,2,2d) }{S_3(2|2,2,2d) }\\ \nonumber
d=k+N
\end{eqnarray}

Next, triple sine function has the product  representation, immediately giving GV representation, plus terms non-perturbative w.r.t. the string coupling constant, see Section \ref{sec:prod}.

We extend this approach to $SO/Sp$ groups in Sections \ref{sect:SO}, \ref{sect:Sp}, \ref{sect:SO/Sp} and obtain similar representation, which now includes  double sine function (i.e. modular quantum dilogarithm function, up to second order polynomial). Particularly, $SO(N)$ partition function is

\begin{eqnarray} \label{SON}
Z=\sqrt{\frac{S_3(2N|2,2,2d)}{S_3(2|2,2,2d)}} \frac{\sqrt{S_2(2N|4,2d)}}{S_2(N|2,2d)}  2^{-\frac{3}{4}}, \\ \nonumber
d=k+N-2
\end{eqnarray}

and $Sp(N)$ partition function is

\begin{eqnarray} \label{SpN}
Z=
\sqrt{\frac{S_3(2N+4|2,2,4d)}{S_3(2|2,2,4d)}} \frac{S_2(N+2|2,4d)}{\sqrt{S_2(2N+4|4,4d)}}2^{-\frac{1}{4}}\\ \nonumber
d=k+N/2+1
\end{eqnarray}

Both formulae lead  to  GV structure form for partition function of  topological strings on orientifolds with non-orientable world sheets included, with odd multicoverings, as obtained in \cite{SV} and \cite{BFM,BFM2}. 

Note the relation between partition functions for classical groups (cf. \cite{OV}) in evident notations:
\begin{eqnarray}
Z(SO(N+1),d)Z(Sp(N-1),\frac{d}{2})=\frac{\sqrt{N}}{2\sqrt{d}}Z(SU(N),d)
\end{eqnarray}

It is an interesting problem to extend the gauge/string duality and geometrical transition to Chern-Simons theory with exceptional groups. Among the aims of present paper is to find a GV type representation for partition function of Chern-Simons theory with exceptional groups, and hence GV integer invariants for corresponding, after geometric transition, manifold. This can be considered as the reasonable step towards establishing a full  duality between corresponding theories.

Idea is the following. There is an interesting  feature of Vogel's results presented in Table \ref{tab:1} that all exceptional groups belong to the same line in Vogel's plane, namely the line $\gamma=2(\alpha+\beta)$. We shall call it an exceptional line, or $Exc$ line, \cite{LM1}. This is the same feature as for classical groups: $SU(N)$ groups occupy the line $\alpha+\beta=0$, $SU(N)$ group corresponds to point $\alpha=-2, \beta=2, \gamma=N$, $SO/Sp$ groups occupy the line $2\alpha+\beta=0$, $SO(N)$ group corresponds to  $\alpha=-2, \beta=4, \gamma=N-4$, $Sp(N)$ corresponds to  $\alpha=-2, \beta=1, \gamma=N/2+2$. Existence of $Exc$ line led Deligne \cite{Del,DM} to a hypotheses that they form a series of Lie algebras, which, roughly formulated, means that they behave uniformly with respect to the decomposition of tensor powers of adjoint representation. In domain of overlapping results of   \cite{Del,DM} coincide with universal formulae restricted to exceptional line, e.g. some dimensions of irreps in powers of adjoint representation coincide with restriction of universal dimension formulae of Landsberg and Manivel \cite{LM1} on exceptional line. 

Having in mind an analogy with classical groups, we consider partition function on exceptional line and obtain in Section \ref{sec:EPF} expression for  partition function in terms of double sine functions, which leads in the same way to GV type representation.  The peculiarity is that we get a sum of two GV representation, i.e. partition function is a product of those of two topological strings, on manifolds with opposite GV integers. One  have to identify geometrically the corresponding Calaby-Yau manifolds, and also some details of representation (such as an  even multicoverings)  still have to be  explained. The answer for Chern-Simons on 3d sphere on exceptional line shows remarkable similarity with classical cases:

\begin{eqnarray} \label{EXC}
Z=    \frac{1}{4\pi\sin\frac{\pi N}{2d}} \sqrt{\frac{2}{N+2}}
  \prod_{p=1}^6 \left( \frac{ S_2(pN|N+2,2d)}{S_2(pN|2,2d) }\right)^{c_p} \\
  d=k+3N
\end{eqnarray}
where $N=n+2$, $n$ from Table \ref{tab:1}, $c_p=1,2,2,2,1,1$ for $p=1,2,3,4,5,6$, respectively. Integers $c_p$ appear to be GV integer invariants.

Of course, an analogy with 't Hooft's 1/N expansion for $SU(N)$ groups is not complete. 't Hooft extended the values of each $SU(N)$ group weight of Feynman digram from integers to an arbitrary $N$ and provide a topological interpretation of each term in $1/N$ expansion of given diagram.  For exceptional line both steps are problematic. 

First step was attempted by Vogel \cite{V0} and Kneissler \cite{Kne} (see also Cvitanovich \cite{Cvitbook,Cvi}) for all simple groups simultaneously. Their motivation was coming from knot theory, but problem they addressed can be formulated directly in terms of field theory diagrams. Namely, consider group weights of vacuum Feynman diagrams of gauge theory without matter. Their vertexes are three-leg, antisymmetric, and set of these diagrams satisfies Jacoby identities, which in this case in knot theory is called IHX relations. Vogel \cite{V0} introduced the operation of  multiplication  of diagrams which turns this set of diagrams (actually Vogel considered diagrams with three external legs, but it is almost the same as vacuum diagrams' algebra, see \cite{CDM}) into an algebra, called $\Lambda$  algebra, and conjecture the set of generators of algebra $\Lambda$. Kneissler studied the consequences of Jacoby relations and under an additional conjecture showed that $\Lambda$ algebra has a character function and is  equivalent to subalgebra of polynomial algebra over three variables. In this way, due to the existence of character function, one would obtain the most general value of each vacuum Feynman diagrams in terms of these three variables. However, Vogel \cite{V0} found an additional relation, which violates abovementioned conjecture, and shows that it leads to his main statement that not all Vasiliev's finite invariants are coming from Lie algebras' weights. From the point of view of our needs this means that there is no natural universal value, i.e. the function of universal parameters, for a given vacuum Feynman diagrams (of sufficiently high order - starting from order of Vogel's additional constraint). Situation is even more complicated since, first, Vogel's conjecture on generators of algebra $\Lambda$  still remains a conjecture and, second, some additional constraints are possible \cite{Pat,V0}. 

It is also worth to mention Deligne's hypothesys \cite{D} on universal characters, which can potentially provide an alternative way for extension of group weights to entire Vogel's plane.  P. Deligne \cite{D}  suggested that universal characters satisfy usual character relations for decomposition of product of representations  at  all values of universal  parameters. This is a natural extension of his earlier idea of series of Lie algebras \cite{Del,DM}. One can try to turn things round and suggest this feature as definition of extension of characters to entire Vogel space. 

From the other side, we actually have a number of universal quantities in gauge theories, as well as in the theory of Lie algebras. They can appear in different way, for example, imagine that some quantity is determined by an equation with parameters completely calculable by low order diagrams. This happen, e.g. for  eigenvalues of higher Casimirs on adjoint representation, universally calculated in \cite{MSV}. In that cases we get a unique universal answer for that quantity.

For a Chern-Simons theory on 3d sphere we have an explicit  universal  expression for partition function in terms of universal character (\ref{gene})  so restricting it on exceptional line one may consider an analog of $1/N$ expansion with some appropriate choice of parameter $N$ on exceptional line. This solve the problem of expressing partition function, (but not separate Feynman diagrams) in terms of some continuous parameter on exceptional line, analog of $N$. The next, most challenging problem of topological interpretation, seems to be much more difficult. However, the  GV form of partition function provides a solution, since in that form it is interpreted as partition function of topological strings on a manifold with given GV integer invariants. 

Altogether, in this way we suggest to extend the gauge/string duality for exceptional line. Few problems still remain to be resolved (see Section \ref{sec:EGV}) - interpretation of all terms in representation obtained, appearance of even multicoverings, identification of Calaby-Yau manifolds with given GV invariants. Actual computations have been carried on for a family of lines $ \gamma=k(\alpha+\beta) $ with integer $k$. It appears that cases $k=1,2$ are distinguished by some cancellations, which allow us to express partition function in terms of Barnes' double gamma functions, further combined into double sine function. The $k=2$ line is exceptional  $Exc$ line.  Line with $k=1$ is an F-line from \cite{Mkr2},  containing groups $ E_6, SO(10)$ at points (-2,6,8), (-2,4,6) respectively. Comparison of answers for $k=1$ and $k=2$ hints on a "building blocks" they are both constructed from, see Sections \ref{sec:EPF}, \ref{sec:EGV}. 

In Section \ref{full} we show that full universal partition function of Chern-Simons theory, at an arbitrary point in Vogel's plane, can be represented as a ratio of quadruple sine functions, which leads to some generalization of GV representation, rising the question whether it corresponds to some "universal" topological string theory. 

In Conclusion we briefly discuss matrix model for exceptional line.

\section{Partition function of SU(N) Chern-Simons theory as a ratio of triple sine functions}

We shall use an expression \cite{M13-2} of partition function of $SU(N)$ theory in terms of  Barnes' multiple gamma functions. 

Barnes' multiple zeta-function \cite{Barnes3}:
 
\begin{eqnarray}\label{zN}
\zeta_N(w,s|a_1,a_2,...,a_N)=\sum_{n_1,...,n_N=0}^{\infty}\frac{1}{(w+a_1n_1+a_2n_2+...+a_N n_N)^s}
\end{eqnarray}
is defined particularly at $\Re w >0, \Re s>N $ and positive parameters $a_i$.

Then Barnes's multiple gamma-functions $\Gamma_N(w)=\Gamma_N(w|a_1,a_2,...,a_N)$ depending on argument $w$ and parameters $a_1, a_2,...,a_N$ are defined in  similarity with definition of  Euler's gamma-function in terms of  Riemann's zeta-function: 
 
\begin{eqnarray}\label{gN}
\ln\Gamma_N(w)=\Psi_N(w)=\Psi_N(w|a_1,a_2,...,a_N)=\\
\partial\zeta_N(w,s|a_1,a_2,...,a_N)|_{s=0}
\end{eqnarray}
This definition follows that of \cite{Rui} and differs from original Barnes' one \cite{Barnes3} by some modular "constant", depending on parameters. It is more convenient for our purposes.

Integral representation  \cite{Rui} of logarithm of multiple gamma function is:

\begin{eqnarray} \label{Psi}
\Psi_N(w)=\int_{0}^{\infty}\frac{dx}{x}\left( e^{-wx} \prod_{j=1}^{N} \frac{1}{(1-e^{-a_jx})} \right.  \\
\left. - x^  {-N}\sum_{n=0}^{N-1}\frac{(-x)^n}{n!}B_{N,n}(w) -\frac{(-1)^N}{N!}e^{-x}B_{N,N}(w)\right) 
\end{eqnarray}
where multiple Bernoulli  polynomials $B_{N,n}(w)$ are defined as:
\begin{eqnarray}
x^{N}e^{-wx}\prod_{j=1}^{N} \frac{1}{(1-e^{-a_jx})}= \sum_{n=0}^{\infty}\frac{(-x)^n}{n!}B_{N,n}(w)
\end{eqnarray}

Both gamma function and multiple Bernoulli polynomials are implied to depend on positive parameters $ a_j $.  Integrals converge provided real part of $w$ is positive. 

We shall need a recurrent relations \cite{Barnes3,Rui} on Barnes' multiple functions. 
These relations have an origin just in  definition (\ref{zN}): if $w=w_0+a_i, \Re w_0>0$, then sum over $n_i$ is effectively starting from $n_i=1$, with zeta function argument being $w_0$. It remains to add and subtract contribution of $n_i=0$ to get a relation: 

\begin{eqnarray}\label{rec}
\zeta_N(w_0+a_i,s|a_1,a_2,...,a_N)=\zeta_N(w_0,s|a_1,a_2,...,a_N)-\\ \nonumber
\zeta_{N-1}(w_0,s|a_1,..., a_{i-1},a_{i+1}...,a_N)
\end{eqnarray}

This straightforwardly translates into recurrence relation on multiple gamma-functions:

\begin{eqnarray}\label{recgam}
\Gamma_N(w_0+a_i|a_1,a_2,...,a_N)=\\
\Gamma_N(w_0|a_1,a_2,...,a_N)/\Gamma_{N-1}(w_0|a_1,..., a_{i-1},a_{i+1}...,a_N)
\end{eqnarray}

Another way of deriving recurrent relations is based on integral representation (\ref{Psi}) and observation in \cite{M13} that if in the linear combination of logarithms of few multiple gamma functions the main terms (i.e. the first terms in (\ref{Psi})) cancel, then all other terms cancel, also, and we have a relation between corresponding multiple gamma functions. As an example (which we shall need later) one can consider identity 

\begin{eqnarray} 
\frac{1}{2}\frac{1}{(1-e^{-2x})^2}+\frac{1}{2}\frac{1}{1-e^{-4x}}=\frac{1}{(1-e^{-2x})(1-e^{-4x})}
\end{eqnarray}
which leads to the  identity between multiple gamma functions:

\begin{eqnarray} \label{gg2}
\sqrt{\Gamma_2(2N|2,2)\Gamma_1(2N|4)}=\Gamma_2(2N|2,4)
\end{eqnarray}

Of course, this can be checked directly, also. Note also that \cite{Rui}:

\begin{eqnarray}\label{G11}
\Gamma_1(w|a)=exp\left( \left( \frac{w}{a}-\frac{1}{2}\right) \ln a \right) \Gamma\left( \frac{w}{a}\right) (2\pi)^{-\frac{1}{2}}
\end{eqnarray}

Particularly:
\begin{eqnarray} 
\Gamma_1(x|x)=\sqrt{\frac{x}{2\pi}} \\
\Gamma_1(x|2x)=\sqrt{\frac{1}{2}}
\end{eqnarray}

By definition 
\begin{eqnarray}
\Gamma_0(w)=1/w
\end{eqnarray}

Partition function of $SU(N)$ Chern-Simons theory on 3d sphere has been expressed  \cite{M13-2} in terms of Barnes' triple and double functions:

\begin{eqnarray}\label{pfgn4}
Z= \frac{\Gamma_3(2(k+N)+4) \Gamma_3(2(k+N)) \Gamma_2(4|2,2)}{\Gamma_3(2k+2)\Gamma_3(2N+2) } \frac{1}{ \sqrt{N}}
\end{eqnarray}
Parameters of triple gamma functions are $(2,2,2(k+N))$.

Let's continue transformations (let $k+N=d$):

\begin{eqnarray}
Z= \frac{\Gamma_3(2d+4) \Gamma_3(2d) \Gamma_2(4|2,2)}{\Gamma_3(2k+2)\Gamma_3(2N+2) } \frac{1}{ \sqrt{N}}=\\
\frac{\Gamma_3(4) \Gamma_3(2d) }{\Gamma_3(2k+2)\Gamma_3(2N+2) } \frac{1}{ \sqrt{N}}=\\
\frac{\Gamma_3(2) \Gamma_3(2d+2) \Gamma_2(2d|2,2d) }{\Gamma_2(2,|2,2d)\Gamma_3(2k+2)\Gamma_3(2N+2) } \frac{1}{ \sqrt{N}}=\\
\frac{1}{ \sqrt{N}} \frac{\Gamma_3(2) \Gamma_3(2d+2)  }{\Gamma_3(2k+2)\Gamma_3(2N+2) } \frac{\Gamma_2(2d+2|2,2d)\Gamma_1(2d|2d)}{\Gamma_2(2,|2,2d)}  =\\
\frac{1}{ \sqrt{N}} \frac{\Gamma_3(2) \Gamma_3(2d+2)  }{\Gamma_3(2k+2)\Gamma_3(2N+2) } \frac{\Gamma_1(2d|2d)}{\Gamma_1(2,|2)}  =\\
\frac{\sqrt{d}}{ \sqrt{N}}  \frac{\Gamma_3(2) \Gamma_3(2d+2)  }{\Gamma_3(2k+2)\Gamma_3(2N+2) } =\\
\frac{\sqrt{d}}{ \sqrt{N}}  \frac{S_3(2k+2) }{S_3(2) }=\frac{\sqrt{d}}{ \sqrt{N}}  \frac{S_3(2N+2) }{S_3(2d+2) }
\end{eqnarray}

Here we used definition of multiple sine functions:

\begin{eqnarray} \label{Sr}
S_r(z|\underline{\omega})=\frac{\Gamma_r(|\omega|-z|\underline{\omega})^{(-1)^r}}{\Gamma_r(z|\underline{\omega})}\\
|\underline{\omega}|=\sum_{j=1}^{r}\omega_j
\end{eqnarray}

Some simple properties of sine functions:
\begin{eqnarray}
S_r(cz|c\underline{\omega})=S_r(z|\underline{\omega})\\
S_r(z+\omega_i|\underline{\omega})=S_r(z|\underline{\omega})S_{r-1}(z|\underline{\omega}_i^-)\\
 \; \underline{\omega}_i^-=(\omega_1,...,\omega_{i-1},\omega_{i+1},...\omega_r)\\
 S_r(z|\underline{\omega})S_r(|\underline{\omega}|-z|\underline{\omega})^{(-1)^r}=1\\
 S_1(z|\omega)=2 \sin\frac{\pi z}{\omega} 
\end{eqnarray}

So, we express partition function as ratio of triple sine functions, up to elementary functions' multiplier. In next Section we use the  product representation of $S_3$. 

\section{Product form of triple sine and geometrical transition in SU(N) Chern-Simons}\label{sec:prod}

According to \cite{Nar,Fad,Fad2} one can present multiple sine functions $S_r, r\geq 2,$ as an exponent of integral over entire real line, bypassing singularity at zero either from upside or downside:

\begin{eqnarray}
S_r(z|\underline{\omega})=\\
\exp\left( (-1)^r \frac{\pi i}{r!}B_{rr}(z|\underline{\omega})+(-1)^r \int_{R+i0}\frac{dx}{x}\frac{e^{zx}}{\prod_{k=1}^r(e^{\omega_i x}-1)} \right)=\\
\exp\left((-1)^{r-1} \frac{\pi i}{r!}B_{rr}(z|\underline{\omega})+(-1)^r \int_{R-i0}\frac{dx}{x}\frac{e^{zx}}{\prod_{k=1}^r(e^{\omega_i x}-1)} \right)
\end{eqnarray}
where $B_{rr}$ are generalized Bernoulli polynomials, for $r=3$ it is:

\begin{eqnarray}
B_{33}(z|w_1,w_2,w_3)=\frac{z^3}{w_1w_2w_3}-\frac{3z^2(w_1+w_2+w_3)}{2w_1w_2w_3}+\\ 
z\frac{w_1^2+w_2^2+w_3^2+3w_1w_2+3w_2w_3+3w_1w_3}{2w_1w_2w_3}-\\
\frac{(w_1+w_2+w_3)(w_1w_2+w_2w_3+w_1w_3)}{4w_1w_2w_3}
\end{eqnarray}

One can close the contour of integration in the first case above in upper, or respectively in lower semiplane and obtain the value of integral in terms of the sum of residues, which gives the product representation for multiple sine. To separate GV contributions from others, we calculate residues at general values of $N$ and $d$:

\begin{eqnarray}
\ln \left( S_3(2N+2|2,2,2d)\right) = \\ 
- \frac{\pi i}{6}B_{33}(2N+2|2,2,2d)+\sum_{n=1}^{\infty}\frac{e^{\tau n} }{n (2\sin\frac{g_s n}{2})^2} +N_1\\
\ln \left( S_3(2|2,2,2d)\right) = - \frac{\pi i}{6}B_{33}(2|2,2,2d)+\sum_{n=1}^{\infty}\frac{1}{n(2\sin\frac{g_s n}{2})^2}+N_2 \\
g_s=\frac{2\pi}{d}, \tau=iNg_s
\end{eqnarray}
where we explicitly wrote contributions coming from zeros of $(e^{2dx}-1)$ in denominators of both sine functions. The remaining terms are 

\begin{eqnarray}
N_1=\sum_{n=1}^{\infty}i\frac{e^{2 i n N \pi } \left(1-2 i n N \pi +e^{2 i d n \pi } (-1-2 i d n \pi +2 i n N \pi )\right)}{2\pi \left(-1+e^{2 i d n \pi }\right)^2 n^2  }\\
N_2=\sum_{n=1}^{\infty}i\frac{1+e^{2 i d n \pi } (-1-2 i d n \pi )}{2\pi \left(-1+e^{2 i d n \pi }\right)^2 n^2  }
\end{eqnarray}
They are non-perturbative w.r.t. the string coupling constant $g_s=2\pi/d$ and have to be compared with known non-perturbative terms in e.g. \cite{PS}. We omit them, as well as Bernoulli polynomials, which contribute into first few terms (up to torus inclusively)  over genus expansion, which are ambiguous. Remaining terms in  partition function of Chern-Simons theory with $SU(N)$ gauge group on 3d sphere are:
\begin{eqnarray} \label{SUNGV}
\ln Z \cong \sum_{n=1}^{\infty}\frac{e^{\tau n} }{n (2\sin\frac{g_s n}{2})^2}-\sum_{n=1}^{\infty}\frac{1}{n(2\sin\frac{g_s n}{2})^2}
\end{eqnarray}
where first term is desired Gopakumar-Vafa structure form of partition function of topological string on resolved conifold \cite{GV,Mar1}, and second one is contribution of constant maps \cite{BCOV,GP,FP}.

\section {Partition function of SO(N) Chern-Simons theory} \label{sect:SO}

In this case we shall not start from general formulae with quadruple Barnes functions, but transform the integrand of integral representation of partition function. This is more powerful method, since provide the implicit use of more identities than just a recurrent relations on Barnes functions. Let's present partition function (\ref{Ftotal}) as product of two multipliers, perturbative one $Z_2$ and non-perturbative $Z_1$:

\begin{eqnarray}
\label{f2}
Z&=&Z_1Z_2  \\
-\ln Z_2&=&\int^{\infty}_0 \frac{dx}{x} \frac{f(x/\delta)-dim}{(e^{x}-1)} \\
-\ln Z_1 &=&(dim/2)\ln(\delta/t) -\int^{\infty}_0 \frac{dx}{x} \frac{f(x/t)-dim}{(e^{x}-1)}
\end{eqnarray}

The key identity is:

\begin{eqnarray}
f(x)=\frac{\text{cosh}[x(N-1)]}{4\text{sinh}\left[\frac{x}{2}\right]^2} -\frac{1}{4\text{sinh}\left[\frac{x}{2}\right]^2}\\ -\frac{\text{sinh}[x(N-1)]}{2\text{sinh}[x]}+\frac{\text{sinh}\left[\frac{x(N-1)}{2}\right]}{\text{sinh}\left[\frac{x}{2}\right]}
\end{eqnarray}
This equation is another form of observation of \cite{SV}, that perturbative part of free energy of $SO(N)$ is the sum of half of that for $SU(N-1)$ plus terms with odd power of $(N-1)$. Indeed, the first line of r.h.s. of this equation is half of character for $SU(N-1)$ group. The non-perturbative part, however is more complicated. 

In analogy with $SU(N)$ we can transform the expression for partition function into a product of multiple gamma functions. 

\begin{eqnarray}
-\ln Z_2= \int_{0}^{\infty} \frac{dx}{x} \frac{1}{\left(e^{d x}-1\right)}\left( \frac{\text{cosh}[x(N-1)]}{4\text{sinh}\left[\frac{x}{2}\right]^2}-\frac{1}{4\text{sinh}\left[\frac{x}{2}\right]^2} \right. \\
\left.  -\frac{\text{sinh}[x(N-1)]}{2\text{sinh}[x]}+\frac{\text{sinh}\left[\frac{x(N-1)}{2}\right]}{\text{sinh}\left[\frac{x}{2}\right]}-\frac{(N-1)N}{2}\right)
\end{eqnarray}
We rescale integration variable $x \rightarrow 2x$ to avoid fractions and then rewrite this as a product of multiple gamma functions. As explained in \cite{M13-2}, if we have the nonsingular at $x=0$ sum of main terms (first terms in integral representation (\ref{Psi}))  of (logarithms of) few multiple gamma functions, we can consider additional terms in their integral representation (\ref{Psi}) as existing ones, since actually they sum up to zero. So e.g. expression above can be written as the sum of logarithms of gamma functions, corresponding to their main terms, entering in expression above. 

\begin{eqnarray}
Z_2=\frac{\Gamma_3(2d+2|2,2,2d)}{\left( \Gamma_3(2d+2-2(N-1))\Gamma_3(2d+2+2(N-1)) \right)^{\frac{1}{2}}} \times  \\
\left(\frac{\Gamma_2(2d+2-2(N-1)|4,2d)}{\Gamma_2(2d+2+2(N-1)|4,2d)} \right)^{\frac{1}{2}} \frac{\Gamma_2(2d+1+(N-1)|2,2d)}{\Gamma_2(2d+1-(N-1)|2,2d)}  \times \\
\left( \Gamma_1(2d|2d) \right)^{\frac{N(N-1)}{2}}
\end{eqnarray}

Expression for $Z_1$ can be obtained from that for $Z_2$ by the fact that $Z=1$ at $k=0$. But we can't consider above terms at $k=0$ separately, unlike to $SU(N)$ case.  The reason is the following: first, we represent character for $SO(N)$ as the sum of terms, part of which coincide with those for $SU(N-1)$ (up to multiplier $1/2$). Then contribution of volume becomes represented in the similar form, but the point is that there $x$ is divided by dual Coxeter number of $SO(N)$, i.e. $N-2$, and not a dual Coxeter number of $SU(N-1)$, which is $N-1$. Correspondingly some integrals become divergent if considered separately. It is seen from e.g. the multiplier $\Gamma_3(2d+2-2(N-1))$ which at $k=0$ get an argument $2d+2-2(N-1)=2(k+N-2)+4-2N=0$. So we have to combine some divergent terms to achieve singularities' cancellation. Then we get for $Z_1$:

\begin{eqnarray}
\frac{f(x)-N(N-1)/2}{e^{2(N-2)x}-1}=  \frac{e^{-2 x}-e^{-2 N x}}{\left(1-e^{-4 x}\right) \left(1-e^{-2 x}\right)}+ \\ \frac{-e^{4 x-3 N x}+e^{2 x-N x}}{\left(1-e^{-2 x}\right) \left(1-e^{2 (2-N) x}\right)}-\frac{e^{-2 (-2+N) x} (-1+N) N}{2 \left(1-e^{-2 (-2+N) x}\right)}
\end{eqnarray}

Note the usual cancellation which decrease the multiplicity of gamma functions in non-perturbative part of partition function. In this case we get double gamma functions instead of triple ones in $Z_2$. Final answer is:

\begin{eqnarray}
Z_1=\frac{\Gamma_2(2,|2,4)}{\Gamma_2(2N|2,4)} \frac{\Gamma_2(N-2|2,2N-4)}{\Gamma_2(3N-4|2,2N-4)} \times  \\
\left( \frac{1}{\Gamma_1(2N-4|2N-4)} \right)^{\frac{N(N-1)}{2}} \left(\frac{N-2}{d} \right)^{\frac{N(N-1)}{4}}
\end{eqnarray}

Next we carry on few transformations, which result in the appearance of triple and double sine functions and a lot of cancellations of uniple gamma functions. First note that

\begin{eqnarray}
\Gamma_3(2d+2|2,2,2d)=\sqrt{\frac{\Gamma_3(2d+2)\Gamma_3(2)}{\Gamma_2(2|2,2)}}=\frac{1}{\sqrt{S_3(2|2,2,2d)\Gamma_2(2|2,2)}}\\
\left( \Gamma_3(2d+2-2(N-1))\Gamma_3(2d+2+2(N-1)) \right)^{\frac{1}{2}}=\\
=\frac{1}{\sqrt{S_3(2N|2,2,2d)\Gamma_2(2N|2,2)}}
\end{eqnarray}

\begin{eqnarray}
\frac{\Gamma_2(2d+2-2(N-1)|4,2d)}{\Gamma_2(2d+2+2(N-1)|4,2d)}=S_2(2N|4,2d) \Gamma_1(2N|4) \\ \frac{\Gamma_2(2d+1+(N-1)|2,2d)}{\Gamma_2(2d+1-(N-1)|2,2d)} =\frac{1}{S_2(N|2,2d)\Gamma_1(N|2)}
\end{eqnarray}

Altogether:

\begin{eqnarray}
Z_2=\sqrt{\frac{S_3(2N|2,2,2d)\Gamma_2(2N|2,2)}{S_3(2|2,2,2d)\Gamma_2(2|2,2)}} \times\\
\frac{\sqrt{S_2(2N|4,2d) \Gamma_1(2N|4)}}{S_2(N|2,2d)\Gamma_1(N|2)}\left( \Gamma_1(2d|2d) \right)^{\frac{N(N-1)}{2}}=\\
\sqrt{\frac{S_3(2N|2,2,2d)}{S_3(2|2,2,2d)}} \frac{\sqrt{S_2(2N|4,2d)}}{S_2(N|2,2d)}  \times\\
 \sqrt{\frac{\Gamma_2(2N|2,2) }{\Gamma_2(2|2,2)}}
    \frac{\sqrt{\Gamma_1(2N|4)}}{\Gamma_1(N|2)} \left( \frac{d}{\pi} \right)^{\frac{N(N-1)}{4}} 
\end{eqnarray}

\begin{eqnarray}
Z_1=\frac{\Gamma_2(2,|2,4)}{\Gamma_2(2N|2,4)} \Gamma_1(N|2)\Gamma_1(N-2|2N-4) \times \\ 
\left( \frac{1}{\Gamma_1(2N-4|2N-4)} \right)^{\frac{N(N-1)}{2}}  \left(\frac{N-2}{d} \right)^{\frac{N(N-1)}{4}}=\\
\frac{\Gamma_2(2,|2,4)\Gamma_1(N|2)}{\Gamma_2(2N|2,4)} \frac{1}{\sqrt{2}} \left( \frac{\pi}{d}  \right)^{\frac{N(N-1)}{4}}
\end{eqnarray}

Total partition function becomes

\begin{eqnarray}
Z=Z_1Z_2=\\
\sqrt{\frac{S_3(2N|2,2,2d)}{S_3(2|2,2,2d)}} \frac{\sqrt{S_2(2N|4,2d)}}{S_2(N|2,2d)}  \times\\
 \sqrt{\frac{\Gamma_2(2N|2,2)\Gamma_1(2N|4) }{\Gamma_2(2|2,2)}} 
\frac{\Gamma_2(2,|2,4)}{\Gamma_2(2N|2,4)} \frac{1}{\sqrt{2}} 
\end{eqnarray}

Now we use an identity \ref{gg2}:
\begin{eqnarray}
\sqrt{\Gamma_2(2N|2,2)\Gamma_1(2N|4)}=\Gamma_2(2N|2,4)
\end{eqnarray}

and its particular case at $N=1$. These identities applied to partition function give final answer
\begin{eqnarray}
Z=\sqrt{\frac{S_3(2N|2,2,2d)}{S_3(2|2,2,2d)}} \frac{\sqrt{S_2(2N|4,2d)}}{S_2(N|2,2d)}  2^{-\frac{3}{4}}
\end{eqnarray}

We see that the same phenomenon happens - all uniple gamma functions cancel and only triple and double sine functions remain. So, the role of group volume multiplier is to remove these uniple gamma functions from perturbative part. This finally leads to beautiful expression above. Its product representation leads (below) to Gopakumar - Vafa type expression, analogous to that of $SU(N)$ case. 

This final expression can be checked directly, at least at $k=0$, as we did. One has to restore some uniple gamma functions which were substituted by their values, and check cancellation of the main terms of multiple gamma functions. 

\section{Partition function of Sp(N) Chern-Simons theory}\label{sect:Sp}

For true symplectic groups $Sp(N)$ $N$ should be even (positive) integer. The dual Coxeter number is then integer $(N/2)+1$. 

One can carry on calculations independently  or use previous results for $SO(N)$ due to $N \rightarrow -N$ duality between $SO(N)$ and $Sp(N)$ characters. This  duality is realized by changing $N\rightarrow -N, k \rightarrow -2k, d \rightarrow -2d, x \rightarrow -x$. Last transformation (of $x$) is implied in integrand $f(x)/(exp(dx)-1)$. Under this transformation an integrand for  $SO(N)$  transforms into that of $Sp(N)$. When some parameters of gamma functions above become negative, for each of them one have to change it sign, add modulus of that parameter to argument, and take an inverse of gamma function. Then we have: 

\begin{eqnarray}
Z_2=\frac{\Gamma_3(4d+2|2,2,4d)}{\left( \Gamma_3(4d-2N)\Gamma_3(4d+2N+4) \right)^{\frac{1}{2}}} \times  \\
\left(\frac{\Gamma_2(4d+2N+4|4,4d)}{\Gamma_2(4d-2N|4,4d)} \right)^{\frac{1}{2}} \frac{\Gamma_2(4d-N|2,4d)}{\Gamma_2(4d+N+2|2,4d)}  \times \\
\left( \Gamma_1(4d|4d) \right)^{\frac{N(N+1)}{2}}
\end{eqnarray}

\begin{eqnarray}
Z_2=\sqrt{\frac{S_3(2N+4|2,2,4d)\Gamma_2(2N+4|2,2)}{S_3(2|2,2,4d)\Gamma_2(2|2,2)}} \times\\
\frac{S_2(N+2|2,4d) \Gamma_1(N+2|2)}{\sqrt{S_2(2N+4|4,4d)\Gamma_1(2N+4|4)}}\left( \Gamma_1(4d|4d) \right)^{\frac{N(N-1)}{2}}=\\
\sqrt{\frac{S_3(2N+4|2,2,4d)}{S_3(2|2,2,4d)}} \frac{S_2(N+2|2,4d)}{\sqrt{S_2(2N+4|4,4d)}}  \times\\
 \sqrt{\frac{\Gamma_2(2N+4|2,2) }{\Gamma_2(2|2,2)}}
    \frac{\Gamma_1(N+2|2)}{\sqrt{\Gamma_1(2N+4|4)}} \left( \frac{2d}{\pi} \right)^{\frac{N(N+1)}{4}} 
\end{eqnarray}

Non-perturbative part is:

\begin{eqnarray}
Z_1=\frac{\Gamma_2(4,|2,4)}{\Gamma_2(2N+6|2,4)} \frac{\Gamma_2(3N+6|2,2N+4)}{\Gamma_2(N+4|2,2N+4)} \times  \\
\left( \frac{1}{\Gamma_1(2N+4|2N+4)} \right)^{\frac{N(N+1)}{2}} \left(\frac{\frac{N}{2}+1}{d} \right)^{\frac{N(N+1)}{4}}=\\
\frac{\Gamma_2(4,|2,4)}{\Gamma_2(2N+6|2,4)} \frac{\Gamma_1(N+2|2N+4)}{\Gamma_1(N+2|2)}=\\
\left( \frac{1}{\Gamma_1(2N+4|2N+4)} \right)^{\frac{N(N+1)}{2}} \left(\frac{\frac{N}{2}+1}{d} \right)^{\frac{N(N+1)}{4}}=\\
\frac{\Gamma_2(4,|2,4)}{\Gamma_2(2N+6|2,4)\Gamma_1(N+2|2)} \left(\frac{\pi}{2d} \right)^{\frac{N(N+1)}{4}} \frac{1}{\sqrt{2}}
\end{eqnarray}

Total partition function:

\begin{eqnarray}
Z=Z_1Z_2=\\
\sqrt{\frac{S_3(2N+4|2,2,4d)}{S_3(2|2,2,4d)}} \frac{S_2(N+2|2,4d)}{\sqrt{S_2(2N+4|4,4d)}}  \times\\
 \sqrt{\frac{\Gamma_2(2N+4|2,2) }{\Gamma_2(2|2,2)}}
\frac{1}{\sqrt{\Gamma_1(2N+4|4)}}  \\
\frac{\Gamma_2(4,|2,4)}{\Gamma_2(2N+6|2,4)}  \frac{1}{\sqrt{2}}=\\
\sqrt{\frac{S_3(2N+4|2,2,4d)}{S_3(2|2,2,4d)}} \frac{S_2(N+2|2,4d)}{\sqrt{S_2(2N+4|4,4d)}}2^{-\frac{1}{4}}
\end{eqnarray}

For comparison we put here the $SO(N)$ partition function from above:

\begin{eqnarray}
Z=\sqrt{\frac{S_3(2N|2,2,2d)}{S_3(2|2,2,2d)}} \frac{\sqrt{S_2(2N|4,2d)}}{S_2(N|2,2d)}  2^{-\frac{3}{4}}
\end{eqnarray}

It is seen that one can be obtained from another by recipe described above. The seeming difference in coefficients also comes from the same recipe: in SO case we have a multiplier in partition function  
\begin{eqnarray}
\Gamma(2|2,4)/\sqrt{\Gamma_2(2|2,2)\Gamma_1(2|4)}=1 
\end{eqnarray}
which transforms into  
\begin{eqnarray}
\frac{\Gamma_2(4|2,4)}{\sqrt{\Gamma_2(2|2,2)\Gamma_1(2|4)}}=\frac{\Gamma_2(2|2,4)}{(\sqrt{\Gamma_2(2|2,2)\Gamma_1(2|4)}\Gamma_1(2|4))}=\sqrt{2} 
\end{eqnarray}

These explicit expressions for partition functions should allow the level-rank duality consideration. However, it is not so straightforward as in $SU(N)$ case and requires further study.

\section{Gopakumar-Vafa representation of SO/Sp Chern-Simons partition functions}\label{sect:SO/Sp}

Now we shall deduce GV representation for Chern-Simons theory on 3d sphere with $SO/Sp$ gauge groups. Partition function for $SO(N)$ is

\begin{eqnarray}
Z=\sqrt{\frac{S_3(2N|2,2,2d)}{S_3(2|2,2,2d)}} \frac{\sqrt{S_2(2N|4,2d)}}{S_2(N|2,2d)}  2^{-\frac{3}{4}}\\
d=k+N-2
\end{eqnarray}

The ratio of triple sine functions was already discussed  in case of $SU(N)$ theory, so we get for their contribution the same answer (\ref{SUNGV}) with $N-1$ instead of $N$ and $(1/2)$ in front. The ratio of double sines gives, in the same approximation (i.e. take into account poles of $1/(e^{2dx}-1)$, only):

\begin{eqnarray}
\ln \sqrt{S_2(2N|4,2d)} = -\sum_{n=1}^{\infty}\frac{ie^{\frac{2\pi  i n(N-1)}{d}}}{4n \: \text{sin}\left[\frac{2  \pi  n}{d}\right]} +...   \\
\ln (S_2(N|2,2d))= -\sum_{n=1}^{\infty}\frac{i e^{\frac{\pi  i n(N-1)}{d}}}{2n \: \text{sin}\left[\frac{  \pi  n}{d}\right]} +...
\end{eqnarray}

In the ratio of double sine functions only odd $n$ contribute: 

\begin{eqnarray}
\ln \frac{\sqrt{S_2(2N|4,2d)}}{S_2(N|2,2d)} = \sum_{n=1,3,5,...} \frac{ie^{\frac{\pi  i n(N-1)}{d}}}{2n  \: \text{sin}\left[\frac{  \pi  n}{d}\right]} +...
\end{eqnarray}

Altogether contribution of these terms into partition function is 

\begin{eqnarray} \label{SOGV}
\ln Z = \sum_{n=1}^{\infty}\frac{e^{\tau n} }{2n \: (2\sin\frac{g_s n}{2})^2} + \sum_{n=1,3,5,...}^{\infty} \frac{e^{\frac{\tau n}{2}}}{2n \: \text{sin}\left(\frac{ g_s n}{2}\right)}-\\ \sum_{n=1}^{\infty}\frac{1}{2n(2\sin\frac{g_s n}{2})^2} +...\\
g_s=\frac{2\pi}{d}, \; \tau=i(N-1)g_s, \;  d=k+N-2
\end{eqnarray}
which is in agreement with \cite{SV}. 

For $Sp(N)$ we have
\begin{eqnarray}
Z=
\sqrt{\frac{S_3(2N+4|2,2,4d)}{S_3(2|2,2,4d)}} \frac{S_2(N+2|2,4d)}{\sqrt{S_2(2N+4|4,4d)}}2^{-\frac{1}{4}}+...
\end{eqnarray}

\begin{eqnarray}
\ln (S_2(N+2|2,4d))= -\sum_{n=1}^{\infty}\frac{i e^{\frac{\pi  i n(N+1)}{2d}}}{2n \: \text{sin}\left[\frac{  \pi  n}{2d}\right]} +... \\
\ln \sqrt{S_2(2N+4|4,4d)} = -\sum_{n=1}^{\infty}\frac{ie^{\frac{\pi  i n(N+1)}{d}}}{4n \: \text{sin}\left[\frac{ \pi  n}{d}\right]} +...
\end{eqnarray}

Ratio of these  functions enters in partition function: 

\begin{eqnarray}
\ln \frac{S_2(N+2|2,4d)}{\sqrt{S_2(2N+4|4,4d)}}=-\sum_{n=1,3,5...}^{\infty}\frac{i e^{\frac{\pi  i n(N+1)}{2d}}}{2n \: \text{sin}\left[\frac{  \pi  n}{2d}\right]}+...
\end{eqnarray}
Together with triple sine contribution:

\begin{eqnarray}
\ln \sqrt{\frac{S_3(2N+4|2,2,4d)}{S_3(2|2,2,4d)}} = \sum_{n=1}^{\infty}\frac{e^{\frac{2\pi i n (N+1)}{2d}} }{2n \: (2\sin\frac{\pi n}{2d})^2}+...
 \end{eqnarray}

we get

 \begin{eqnarray}
\ln Z = \sum_{n=1}^{\infty}\frac{e^{\frac{2\pi i n (N+1)}{2d}} }{2n \: (2\sin\frac{\pi n}{2d})^2}  -\sum_{n=1,3,5...}^{\infty}\frac{i e^{\frac{\pi  i n(N+1)}{2d}}}{2n \: \text{sin}\left[\frac{  \pi  n}{2d}\right]}+...=\\  \label{SpGV}
\sum_{n=1}^{\infty}\frac{e^{\tau n}}{2n \: (2\sin\frac{g_s n}{2})^2}  -\sum_{n=1,3,5...}^{\infty}\frac{i e^{\frac{\tau n}{2}}}{2n \: \text{sin}\left[\frac{  g_s  n}{2}\right]}-\\ 
\sum_{n=1}^{\infty}\frac{1}{2n(2\sin\frac{g_s n}{2})^2} +...\\ 
g_s=\frac{2\pi}{2d}, \; \tau=\frac{2\pi i (N+1)}{2d}, \; d=k+\frac{N}{2}+1
 \end{eqnarray}
 
where we introduce string coupling with additional $(1/2)$ multiplier (this multiplier is often missed in literature). Comparing (\ref{SOGV}) for $SO(N)$ and (\ref{SpGV}) for $Sp(N)$, and identifying string coupling and Kahler parameters as above we get the well-known result that the  only difference between $SO$ and $Sp$ cases is in the sign of contribution of non-orientable worldsheets.

\section{Universal character on exceptional line}

In calculation of partition function according to (\ref{Ftotal}) the main element is universal character. We  consider it on  lines $ \gamma=k(\alpha+\beta) $ with natural $k$. $k=2$ corresponds to exceptional line Exc.

Parameterize these lines as $\alpha=z<0, \beta=1-z>0, \gamma = k(\alpha+\beta)=k>0$. There is a key identity:
\begin{eqnarray}
\text{f}(x)=\text{X}[x,k]Z(x)+\text{Y}[x,k],\\
Z(x)=\frac{1}{-1+e^{x z/2}}+\frac{1}{-1+e^{x (1-z)/2}} ,\\
\text{X}[\text{x},\text{k}]\text{=} -\text{csch}[k x/4]\text{csch}[x/4]\text{  }\text{sinh}[(2+k) x/4] \times\\ 
\frac{1}{2} \left(-e^{-x/4}-e^{x/4}+e^{-(3+4 k) x/4}+e^{(3+4 k) x/4}\right),\\
\text{Y}[\text{x},\text{k}]\text{=}-\text{csch}[k x/4]\text{csch}[x /4] \text{sinh}[(2+k) x/4]\\
\frac{1}{2} \left(-2 e^{-x/4}+e^{-(3+4 k) x/4}+e^{(3+4 k) x/4}\right)
\end{eqnarray}

Cases $k=1,2$ are distinguished by the fact that for these $k$ $X[x,k]$ and $Y[x,k]$ are the sums of exponents, and $X[x/t,k]$ is divisible on $(-1+e^x)$.
The function $f(x)$ is even, which is reflected in this representation by the following features of $X$ and $Y$:

\begin{eqnarray}
Z(-x)=-Z(x)-2\\
\text{X}[-x,k]+\text{X}[x,k]=0\\
\text{Y}[-x,k]-\text{Y}[x,k]+2\text{X}[x,k]=0
\end{eqnarray}

The values of $X$ and $Y$  at $k=1,2$ are following.

$k=1$:
\begin{eqnarray}
X[x,1]=e^{-2 x}+2 e^{-3 x/2}+3 e^{-x}+  \\ 
2 e^{-x/2}-2 e^{x/2}-3 e^x-2 e^{3 x/2}-e^{2 x}\\
\text{X}[x/2,1]\frac{1}{-1+e^{x }}=-e^{-x} \left(1+e^{x/4}+e^{x/2}\right)^2=\\
-1-e^{-x}-2 e^{-3 x/4}-3 e^{-x/2}-2 e^{-x/4}\\
\text{Y}[x,1]=
-1+e^{-2 x}+2 e^{-3 x/2}+3 e^{-x}+  \\ e^{-x/2}-3 e^{x/2}-3 e^x-2 e^{3 x/2}-e^{2 x}
\end{eqnarray}

$k=2$:

\begin{eqnarray}
\text{X}[x,2]=
e^{-3 x}+e^{-5 x/2}+2 e^{-2 x}+2 e^{-3 x/2}+2 e^{-x}+  \\  
e^{-x/2}-e^{x/2}-2 e^x-2 e^{3 x/2}-2 e^{2 x}-e^{5 x/2}-e^{3 x}\\
\text{X}[x/3,2]\frac{1}{-1+e^{x }}=  \\
-1-e^{-x}-e^{-5 x/6}-2 e^{-2 x/3}-2 e^{-x/2}-2 e^{-x/3}-e^{-x/6}\\
\text{Y}[x,2]=
e^{-3 x}+e^{-5 x/2}+2 e^{-2 x}+2 e^{-3 x/2}+ \\ 
2 e^{-x}-2 e^{x/2}-2 e^x-2 e^{3 x/2}-2 e^{2 x}-e^{5 x/2}-e^{3 x}
\end{eqnarray}

\section{Partition function of Chern-Simons theory on exceptional line as ratio of double sine functions}  \label{sec:EPF}

For connection with double sine function  the antisymmetry of $X(x)$ is important.
We write 

\begin{eqnarray}
X[x,k]=\sum_p c_p (e^{-\frac{p}{2} x}-e^{\frac{p}{2} x})\\
Y[x,k]=-\delta_{k,1}-(e^{-\frac{1}{2} x}+e^{\frac{1}{2} x}) +\sum_p c_p (e^{-\frac{p}{2} x}-e^{\frac{p}{2} x})
\end{eqnarray}
with integer $c_p$ and p=1,...2(k+1).

For $k=1$ $c_p=2,3,2,1$ , for $p=1,2,3,4$, respectively, and for $k=2$  $c_p=1,2,2,2,1,1$, for $p=1,...,6$, respectively. Note that we use the same notation for two different sequences. Hopefully, it is easy to identify what case - $k=1$ or $k=2$ - we are dealing with. 

We have to put expression for $f(x)$ into  expression for partition function (\ref{Ftotal}). Since for $k=1,2$ $X(x/t,k)$ is divisible on $(-1+e^x)$, the double gamma functions will appear  from the $X(x/\delta)$  terms, only.

So, "typical" term from $X$ is 

\begin{eqnarray}
c_p (e^{-p x/2\delta}-e^{p x/2\delta})\frac{1}{(-1+e^{z x/2\delta})(-1+e^{x})}
\end{eqnarray}

As on classical lines, each such term, considered to be integrated over $x$ from $0$ to infinity, give rise to main term in integral representations \cite{Rui}  of logarithms of Barnes' multiple gamma function. "Multiple" here corresponds to double and uniple.

The  gamma functions corresponding to typical term of $X$  above are (one minus is coming from denominators, another one from connection between free energy and partition function):  

\begin{eqnarray}
\left( \frac{\Gamma_2(\frac{p}{2\delta}+1|1,-\frac{z}{2\delta})}{\Gamma_2(-\frac{p}{2\delta}+1|1,-\frac{z}{2\delta})}\right)^{c_p}
\end{eqnarray}
We assume $\delta$ large enough so that  $1>p/2\delta$.

Next, use the recurrent relations for gamma functions:

\begin{eqnarray}
\Gamma_2(\frac{p}{2\delta}+1|1,-\frac{z}{2\delta})=\Gamma_2(\frac{p}{2\delta}|1,-\frac{z}{2\delta})/\Gamma_1(\frac{p}{2\delta}|-\frac{z}{2\delta})\\
\Gamma_2(-\frac{p}{2\delta}+1|1,-\frac{z}{2\delta})=\Gamma_2(-\frac{p}{2\delta}+1-\frac{z}{2\delta}|1,-\frac{z}{2\delta})\Gamma_1(-\frac{p}{2\delta}+1|1)
\end{eqnarray}

and get instead of above:

\begin{eqnarray}
\left( \frac{\Gamma_2(\frac{p}{2\delta}|1,-\frac{z}{2\delta})}{\Gamma_2(-\frac{p}{2\delta}+1-\frac{z}{2\delta}|1,-\frac{z}{2\delta})\Gamma_1(-\frac{p}{2\delta}+1|1)\Gamma_1(\frac{p}{2\delta}|-\frac{z}{2\delta})}\right)^{c_p}=\\ \label{S2-1}
\left( \frac{1}{S_2(\frac{p}{2\delta}|1,-\frac{z}{2\delta})\Gamma_1(-\frac{p}{2\delta}+1|1)\Gamma_1(\frac{p}{2\delta}|-\frac{z}{2\delta})}\right)^{c_p}
\end{eqnarray}

i.e. we get a double sine function (\ref{Sr}) instead of two double gamma function. 

Another term in product is coming from the second term with $X$, i.e. with $y=(1-z)$ instead of $z$. Since $y$ is positive at negative $z$, one have to multiply on $\exp((1-z)x/2\delta)$, and also have one minus sign less. Then contribution is:

\begin{eqnarray}
c_p (e^{-\frac{px}{2\delta}}-e^{\frac{px}{2\delta}})\frac{1}{(-1+e^{\frac{yx}{2\delta}})(-1+e^{x})}
\end{eqnarray}

\begin{eqnarray}
\left( \frac{\Gamma_2(-\frac{p}{2\delta}+\frac{y}{2\delta}+1|1,\frac{y}{2\delta}) }{  \Gamma_2(\frac{p}{2\delta}+\frac{y}{2\delta}+1|1,\frac{y}{2\delta})  }\right)^{c_p}
\end{eqnarray}

By recurrence relations:
\begin{eqnarray}
 \Gamma_2(\frac{p}{2\delta}+\frac{y}{2\delta}+1|1,\frac{y}{2\delta}) =\frac{ \Gamma_2(\frac{p}{2\delta}|1,\frac{y}{2\delta})\Gamma_0(\frac{p}{2\delta})}{\Gamma_1(\frac{p}{2\delta}|\frac{y}{2\delta})\Gamma_1(\frac{p}{2\delta}|1)}
\end{eqnarray}

we get for this contribution

\begin{eqnarray}
\left( \frac{\Gamma_2(-\frac{p}{2\delta}+\frac{y}{2\delta}+1|1,\frac{y}{2\delta}) }{  \Gamma_2(\frac{p}{2\delta}+\frac{y}{2\delta}+1|1,\frac{y}{2\delta})  }\right)^{c_p}= \\ \label{S2-2}
\left( \frac{ S_2(\frac{p}{2\delta}|1,\frac{y}{2\delta})\Gamma_1(\frac{p}{2\delta}|\frac{y}{2\delta})\Gamma_1(\frac{p}{2\delta}|1)}{\Gamma_0(\frac{p}{2\delta})}\right)^{c_p}
\end{eqnarray}

where 
\begin{eqnarray}
\Gamma_0(w)=1/w
\end{eqnarray}
So, double sine multiplier of partition function is:

\begin{eqnarray}
A_0=\prod_p \left( \frac{ S_2(\frac{p}{2\delta}|1,\frac{y}{2\delta})}{S_2(\frac{p}{2\delta}|1,-\frac{z}{2\delta}) }\right)^{c_p}
\end{eqnarray}

Remaining (from above) multiplyers (mainly uniple gamma functions) are $A_1$ from (\ref{S2-1}) and $A_2$ from (\ref{S2-2})

\begin{eqnarray}
A_1=\prod_p \left( \frac{\Gamma_1(\frac{p}{2\delta}|\frac{y}{2\delta})\Gamma_1(\frac{p}{2\delta}|1)}{\Gamma_0(\frac{p}{2\delta})}\right)^{c_p}  \\
A_2=\prod_p \left( \frac{1}{\Gamma_1(-\frac{p}{2\delta}+1|1)\Gamma_1(\frac{p}{2\delta}|-\frac{z}{2\delta})}\right)^{c_p}
\end{eqnarray}
Slightly regrouping multipliers, introduce $B_1, B_2$ instead of $A_1, A_2$ :

\begin{eqnarray}
B_1=\prod_p \left( \frac{\Gamma_1(\frac{p}{2\delta}|\frac{y}{2\delta})}{\Gamma_1(\frac{p}{2\delta}|-\frac{z}{2\delta})}\right)^{c_p}  \\
B_2=\prod_p \left( \frac{\Gamma_1(\frac{p}{2\delta}|1)}{\Gamma_1(-\frac{p}{2\delta}+1|1)\Gamma_0(\frac{p}{2\delta})}\right)^{c_p} \\
B_1B_2=A_1A_2
\end{eqnarray}

Next let's combine all $ \Gamma_1 $ functions, appearing both from $X$ (above and from $f(x/t)$) and $Y$, see below, sources. We also shall explicitly note minus signs, appearing from different sources, to make easier follow calculations. 

Terms from $Y$ from terms $f(x/\delta)$ are (one minus is coming from definition of free energy, no other minus signs):

\begin{eqnarray}
A_3=\Gamma_1(1|1)^{\delta_{k,1}}\Gamma_1(\frac{1}{2\delta}+1|1)\Gamma_1(-\frac{1}{2\delta}+1|1) \times \\
\prod_p \left(\frac{\Gamma_1(-\frac{p}{2\delta}+1|1)}{\Gamma_1(\frac{p}{2\delta}+1|1)} \right)^{c_p}
\end{eqnarray}

Next we have to take into account terms coming from $f(x/t)$. We write the key identity for $f(x/t)$ and divide on $(-1+e^x)$. Note that $t=k+1$ on our lines.

\begin{eqnarray}
\frac{\text{f}[x/(k+1),z,k]}{-1+e^x}= \\ 
 \frac{\text{X}[x/(k+1),k]}{-1+e^x}\left(\frac{1}{-1+e^{x z/2(k+1)}}+\frac{1}{-1+e^{x (1-z)/2(k+1)}}\right)+\\ 
 \frac{\text{Y}[x/(k+1),k]}{-1+e^x}
\end{eqnarray}

Here terms $X$ and $Y$ are not convergent, under integral sign,  at upper limit separately. We add to this the l.h.s. of identity below:
\begin{eqnarray}
\frac{1}{-1+e^{z x/2(k+1)}}+1-\frac{e^{z x/2(k+1)}}{-1+e^{z x/2(k+1)}}=0
\end{eqnarray}

and combine terms as:
\begin{eqnarray}
\left( \frac{\text{X}[x/(k+1),k]}{-1+e^x}+1\right) \frac{1}{-1+e^{x z/2(k+1)}}+\\
\frac{\text{X}[x/(k+1),k]}{-1+e^x} \frac{1}{-1+e^{x (1-z)/2(k+1)}}+ \\
\left( \frac{\text{Y}[x/(k+1),k]}{-1+e^x}+1 \right)-\frac{e^{z x/2(k+1)}}{-1+e^{z x/2(k+1)}}
\end{eqnarray}

Now all four terms are convergent on upper limit, and we can transform them separately into gamma functions.
Note that the  ratio $ \text{X}[x/(k+1),k]/(-1+e^x)$ is equal to (for $k=1,2$):

\begin{eqnarray}
\text{X}[x/(k+1),k]/(-1+e^x)=-1-\sum_p c_p e^{-\frac{px}{2(k+1)}}
\end{eqnarray}
We add $1$ to this expression and ensure convergence of integral, which appears to be equal to (three minus signs appear - from denominator, minus in front of $f(x/t)$ and from definition of free energy):

\begin{eqnarray}
A_4=\prod_{p} \left( \Gamma_1\left( \frac{p}{2(k+1)}| \frac{-z}{2(k+1)}\right) \right)^{c_p}
\end{eqnarray}

Second term with $X$ gives (two minuses, since one minus less, from denominator): 

\begin{eqnarray}
A_5=\frac{1}{\Gamma_1(\frac{(1-z)}{2(k+1)}|\frac{(1-z)}{2(k+1)})}
\prod_p \left(\frac{1}{\Gamma_1(\frac{p}{2(k+1)}+\frac{(1-z)}{2(k+1)}|\frac{(1-z)}{2(k+1)})}\right)^{c_p}
\end{eqnarray}

Third term, with $Y$,  (two minuses)):

\begin{eqnarray}
A_6= \frac{\Gamma_1(2|1)}{\Gamma_1(1|1)^{\delta_{k,1}}} \left(\frac{1}{\Gamma_1(1-\frac{1}{2(k+1)}|1)\Gamma_1(1+\frac{1}{2(k+1)}|1)} \right) \\
 \prod_{p=1}^{2k+1} \left(\frac{ \Gamma_1(1+\frac{p}{2(k+1)}|1)}{\Gamma_1(1-\frac{p}{2(k+1)}|1)} \right)^{c_p}
\end{eqnarray}
Here we particularly use $c_{2(k+1)}=1$, both for $k=1$ and $k=2$.

Finally, the last, fourth term receives minus from free energy definition, has minus in front, minus from denominator and minus in front of $f$: 

\begin{eqnarray}
A_7=\Gamma_1(-\frac{z}{2(k+1)}|-\frac{z}{2(k+1)})
\end{eqnarray}

Now we shall combine all multipliers. 
A lot of cancellations take place in product $B_2A_3$:

\begin{eqnarray}
B_2A_3=\Gamma_1(1|1)\Gamma_1(1-\frac{1}{2\delta}|1)\Gamma_1(1+\frac{1}{2\delta}|1)=\\
\frac{1}{4\sqrt{2\pi}\delta\sin\frac{\pi}{2\delta}}
\end{eqnarray}
At last line we use  (\ref{G11}). 
Next consider product of $B_1, A_4$ and $A_5$. We  use (\ref{G11}) again and get:

\begin{eqnarray}
B_1A_4A_5=
\sqrt{2\pi\frac{2(k+1)}{y}}
\prod_{p} \left(\Gamma_0(\frac{p}{2(k+1)})\right)^{c_p}
\left(\frac{k+1}{\delta}\right)^{\frac{\sum pc_p}{yz}} 
\end{eqnarray}

Next, $A_7$ is:

\begin{eqnarray}
A_7=\Gamma_1(-\frac{z}{2(k+1)}|-\frac{z}{2(k+1)})=\sqrt{\frac{-z}{2(k+1) 2\pi}}
\end{eqnarray}

So, product $B_1A_4A_5A_7$ is:

\begin{eqnarray}
B_1A_4A_5A_7=
\sqrt{\frac{-z}{y}}
\prod_{p} \left(\Gamma_0(\frac{p}{2(k+1)})\right)^{c_p}
\left(\frac{k+1}{\delta}\right)^{\frac{\sum pc_p}{yz}} 
\end{eqnarray}
where  $\sum_p p c_p=18$ for $k=1$ and  $\sum_p p c_p=30$ for $k=2$.

So, interestingly, all uniple gamma functions cancel, only double sine functions remain, up to explicit elementary functions of $z, \delta$, and numerical multipliers. The final answer for partition function is:
\begin{eqnarray}
Z=  \left( \frac{k+1}{\delta} \right)^{\frac{dim}{2}}  A_0A_1A_2A_3A_4A_5A_6A_7=\\
   \left( \frac{k+1}{\delta} \right)^{\frac{dim}{2}}A_0B_1B_2A_3A_4A_5A_6A_7=\\
\left( \frac{k+1}{\delta} \right)^{\frac{dim}{2}}   \prod_p \left( \frac{ S_2(\frac{p}{2\delta}|1,\frac{y}{2\delta})}{S_2(\frac{p}{2\delta}|1,-\frac{z}{2\delta}) }\right)^{c_p}  \times \frac{1}{4\sqrt{2\pi}\delta\sin\frac{\pi}{2\delta}} \times \\
\sqrt{\frac{-z}{y}}
\prod_{p} \left(\Gamma_0(\frac{p}{2(k+1)})\right)^{c_p}
\left(\frac{k+1}{\delta}\right)^{\frac{\sum pc_p}{yz}}  A_6=\\
\left( \frac{k+1}{\delta} \right)^{-\frac{2+k}{2k}}       
\frac{1}{\delta\sin\frac{\pi}{2\delta}} \sqrt{\frac{-z}{y}}
 \prod_p \left( \frac{ S_2(\frac{p}{2\delta}|1,\frac{y}{2\delta})}{S_2(\frac{p}{2\delta}|1,-\frac{z}{2\delta}) }\right)^{c_p}  Q \\  
Q= \frac{1}{4\sqrt{2\pi}} \prod_{p} \left(\Gamma_0(\frac{p}{2(k+1)})\right)^{c_p}  A_6
\end{eqnarray}

where finally we separate numerical coefficient $Q$ and parameter-dependent part.  

Calculating numerical coefficient, for exceptional line ($k=2$) answer is 

\begin{eqnarray}
Z=   \frac{1}{4\pi\sin\frac{\pi}{2\delta}} \sqrt{\frac{-z}{y}}
 \prod_p \left( \frac{ S_2(\frac{p}{2\delta}|1,\frac{y}{2\delta})}{S_2(\frac{p}{2\delta}|1,-\frac{z}{2\delta}) }\right)^{c_p} 
\end{eqnarray}
For $k=1$ line answer is:
\begin{eqnarray}
Z=   \frac{\sqrt{\delta}}{2\sqrt{2\pi}\sin\frac{\pi}{2\delta}} \sqrt{\frac{-z}{y}}
 \prod_p \left( \frac{ S_2(\frac{p}{2\delta}|1,\frac{y}{2\delta})}{S_2(\frac{p}{2\delta}|1,-\frac{z}{2\delta}) }\right)^{c_p} 
\end{eqnarray}

Comparison of $k=1$ and $k=2$ cases hints that they are constructed from the building blocks given above at $p=1,2,...$

\section{Gopakumar-Vafa form of Chern-Simons theory on exceptional line.} \label{sec:EGV}

Following \cite{Nar} (see also \cite{Fad,Fad2,Vol,Kash}) we use the integral representation for (logarithm of) the double sine as an integral over entire real axis, with pole at zero point bypassing from above, and next close the contour of integration in upper semiplane and obtain a product representation of double sine. 

Considered multiplier of partition function is 
\begin{eqnarray}
A_0=\prod_p \left( \frac{ S_2(\frac{p}{2\delta}|1,\frac{y}{2\delta})}{S_2(\frac{p}{2\delta}|1,\frac{-z}{2\delta}) }\right)^{c_p}
\end{eqnarray}

According to \cite{Nar}  

\begin{eqnarray}
S_2(a|b_1,b_2)=exp\left(  \frac{\pi i}{2}B_{22}(a|b_1,b_2) +\int_{R+i0}\frac{dx}{x} \frac{e^{ax}}{(e^{b_1x}-1)(e^{b_2x}-1)}  \right) \\
B_{22}(a|b_1,b_2)= \frac{a^2}{b_1b_2}-a\frac{b_1+b_2}{b_1b_2}+\frac{b_1^2+b_2^2+3b_1b_2}{6b_1b_2}
\end{eqnarray}

Evaluating integral by closing contour in upper semiplane, we get (take already $b_1=1, b_2=b>0$)

\begin{eqnarray}
\int_{R+i0}\frac{dx}{x} \frac{e^{ax}}{(e^x-1)(e^{bx}-1)} = \sum_{n=1}^{\infty} \frac{1}{n} \left(  \frac{e^{2\pi i n a}}{e^{2\pi inb}-1} +  \frac{e^{2\pi i n a/b}}{e^{2\pi in/b}-1} \right)
\end{eqnarray}

Consequently, for two multipliers in $A_0$ we get contributions into free energy
\begin{eqnarray} \label{a01}
\sum_p \sum_{n=1}^{\infty} \frac{c_p}{n} \left(  \frac{e^{2\pi i n \frac{p}{2\delta}}}{e^{2\pi in\frac{-z}{2\delta}}-1} +  \frac{e^{2\pi i n \frac{p}{-z}}}{e^{2\pi in\frac{2\delta}{-z}}-1} \right)
\end{eqnarray}

and 

\begin{eqnarray} \label{a02}
-\sum_p \sum_{n=1}^{\infty} \frac{c_p}{n} \left(  \frac{e^{2\pi i n \frac{p}{2\delta}}}{e^{2\pi in\frac{1-z}{2\delta}}-1} +  \frac{e^{2\pi i n \frac{p}{1-z}}}{e^{2\pi in\frac{2\delta}{1-z}}-1} \right)
\end{eqnarray}

Let's introduce the string coupling constant according to usual formula 

\begin{eqnarray}
g_s=\frac{2\pi}{d}
\end{eqnarray}
where $d$ is $\delta$ in minimal normalization, when the only negative one of Vogel's parameter's is equal to -2, see Table \ref{tab:1}. Connection of our parametrization with minimal one can be established by multiplying former on $(-2/z)$:  

\begin{eqnarray}
(\alpha,\beta,\gamma, \delta)=(-2,-2(1-z)/z, -2k/z,-2\delta/z ) \\
d=(-2\delta/z)
\end{eqnarray}
Then
\begin{eqnarray} \label{Ecoup1}
g_s=-\frac{\pi z}{\delta}
\end{eqnarray}

Put this into the first  term in (\ref{a01}): 

\begin{eqnarray} \label{Estring}
\sum_p \sum_{n=1}^{\infty} \frac{c_p}{2ni}   \frac{e^{\frac{n\tau_p}{2}}}{\sin(\frac{ng_s}{2})} \\
\tau_p=-i g_s (1+\frac{2p}{z})
\end{eqnarray}
or, introducing parametrization of our two lines $(-2,N+2,kN), k=1,2$ and correspondingly $z=-2/N$ :

\begin{eqnarray} 
\sum_p \sum_{n=1}^{\infty} \frac{c_p}{2ni}   \frac{e^{\frac{n\tau_p}{2}}}{\sin(\frac{ng_s}{2})} \\
 \tau_p=i g_s (pN-1)
\end{eqnarray}

This has the form of contribution of non-orientable surfaces into free energy of topological strings  with coupling $g_s$, $\tau_p$ values of parameters of manifold after geometric transition, Gopakumar-Vafa integer invariant's values $c_p$ and other invariants equal to zero. See, e.g. \cite{SV,BFM,BFM2}. The only difference is that here the sum is over all positive values of $n$, while in \cite{SV,BFM,BFM2} sum is over odd positive values of $n$. The absence of terms (of type $1/sin^2$) with Euler characteristics of manifold  can signal that it is zero. 

However, there is another contribution of poles of $1/(e^x-1)$, given in the first term in (\ref{a02}). It coincides, up to the sign, with  (\ref{Estring}), provided we change $(-z)$ on $(1-z)$, i.e. introduce another string coupling and parameters of  Calaby-Yau manifold:
  
\begin{eqnarray}  \label{Estring2}
-\sum_p \sum_{n=1}^{\infty} \frac{c_p}{2ni}   \frac{e^{\frac{n \tilde{\tau}_p}{2}}}{\sin(\frac{n \tilde{g}_s}{2})} \\
\tilde{g}_s=\frac{\pi (1-z)}{\delta},  \; \;  \tilde{\tau}_p=-i \tilde{ g}_s (1-\frac{2p}{1-z})
\end{eqnarray}

According to this one can assume that partition function on $Exc$ line (and similarly on F line) is the product of two partition functions of two strings on two Calaby-Yau manifolds, with opposite GV invariants and Euler characteristics, and different parameters, given above.  Otherwise, leaving one string, one should be able to present additional terms (\ref{Estring2}) in GV form with initial string coupling (\ref{Ecoup1}). This problem requires further investigation, particularly, string representation of other quantities, besides partition function, should be studied.

\section{Universal Chern-Simons partition function as ratio of quadruple sine functions. } \label{full}

Consider full universal partition function of Chern-Simons theory on 3d sphere, expressed in terms of quadruple Barnes' gamma functions \cite{M13-2}:
\begin{eqnarray}\label{tpfgn}
Z= \frac{\Gamma_4(w_1) \Gamma_4(w_2) \Gamma_4(w_3) \Gamma_4(w_7)  }{\Gamma_4(w_4) \Gamma_4(w_5) \Gamma_4(w_6) \Gamma_4(w_8) }
  \frac{\Gamma_4(v_4) \Gamma_4(v_5) \Gamma_4(v_6) \Gamma_4(v_8)  }{\Gamma_4(v_1) \Gamma_4(v_2) \Gamma_4(v_3) \Gamma_4(v_7) }
 \end{eqnarray}
where (positive) parameters of gamma functions with  arguments $w_i$ are $-\alpha,\beta,\gamma, 2\delta$, those for gamma functions with arguments $v_i$ are $-\alpha,\beta,\gamma, 2t$,  arguments  $w_i$ and $v_i=w_i|_{\delta =t}$ are:

\begin{eqnarray}\label{ws}
w_1&=& 2\delta-2\alpha, \\
w_2&=& 2\delta-\alpha-\beta, \\
w_3&=& 2\delta-\alpha-\gamma,\\
w_4&=& 2\delta+\alpha+\beta+\gamma,\\
w_5&=& 2\delta+2\beta+\gamma,\\
w_6&=& 2\delta+\beta+2\gamma,\\
w_7&=& 2\delta+2\alpha+3\beta+3\gamma,    \\ \label{ws2}
w_8&=& 2\delta-3\alpha-2\beta-2\gamma,
\end{eqnarray}
  
\begin{eqnarray}\label{vs}
v_1&=& 2t-2\alpha,\\
v_2&=& t+\gamma,\\
v_3&=& t+\beta,\\
v_4&=& 3t,\\
v_5&=& 2t+2\beta+\gamma,\\
v_6&=& 2t+\beta+2\gamma,\\
v_7&=& 5t-\alpha,    \\ \label{vs2}
v_8&=& -\alpha,
\end{eqnarray}

There are relations 

\begin{eqnarray}
\frac{\Gamma_4(w_1)}{\Gamma_4(w_4)}&=&\frac{\Gamma_4(w_1)\Gamma_3(\alpha+\beta+\gamma|-\alpha,\beta,\gamma)}{\Gamma_4(\alpha+\beta+\gamma)}=\\
&=& S_4(\alpha+\beta+\gamma)\Gamma_3(\alpha+\beta+\gamma)\\
\frac{\Gamma_4(w_2)}{\Gamma_4(w_5)}&=& S_4(2\beta+\gamma)\Gamma_3(2\beta+\gamma)\\
\frac{\Gamma_4(w_3)}{\Gamma_4(w_6)}&=& S_4(\beta+2\gamma)\Gamma_3(\beta+2\gamma)\\
\frac{\Gamma_4(w_7)}{\Gamma_4(w_8)}&=& \left(S_4(2\alpha+3\beta+3\gamma)\Gamma_3(2\alpha+3\beta+3\gamma) \right)^{-1}
\end{eqnarray}

Same relations are valid with $\delta \rightarrow t$. When inserting them into expression for $Z$ all triple gamma functions cancel and we get:
\begin{tiny}

\begin{eqnarray}
Z=\frac{S_4(\alpha+\beta+\gamma|-\alpha,\beta,\gamma,2\delta)S_4(2\beta+\gamma|-\alpha,\beta,\gamma,2\delta)S_4(\beta+2\gamma|-\alpha,\beta,\gamma,2\delta)}{S_4(2\alpha+3\beta+3\gamma|-\alpha,\beta,\gamma,2\delta)} \times \\
\frac{S_4(2\alpha+3\beta+3\gamma|-\alpha,\beta,\gamma,2t)}{S_4(\alpha+\beta+\gamma|-\alpha,\beta,\gamma,2t)S_4(2\beta+\gamma|-\alpha,\beta,\gamma,2t)S_4(\beta+2\gamma|-\alpha,\beta,\gamma,2t)}
\end{eqnarray}

\end{tiny}

This representation differs qualitatively from what we have for classical and exceptional lines in previous sections, since second fraction doesn't depend from coupling constant and depends from groups, only. Nevertheless, we assume that it is possible, both for classical and exceptional lines, directly transform this expression into those given above. 

For an arbitrary point in Vogel's plane this expression can be further transformed into GV type representation, which, however, will have cube of sine functions in denominator, i.e. will differ from original GV structure. It is an open question whether that formulae are interpretable as some "universal topological string". 

\section{Conclusion}\label{concl}

We show, that partition function of Chern-Simons theory on 3d sphere for different gauge groups can be expressed in terms of multiple sine function. It follows from universal expression for partition function derived in \cite{M13} as a function of Vogel's universal parameters. Exact expressions for classical lines (i.e. for gauge groups $SU(N), SO(N), Sp(N)$ with arbitrary $N$)  allow further discussion of their relations, level-rank duality, etc. Most important feature or representation in terms of multiple sine functions is an immediate derivation of Gopakumar-Vafa form of partition function and hence a gauge/string duality and corresponding geometrical transition. The similar representation of partition function  on an exceptional line (the line in Vogel's plane containing exceptional groups) in terms of double sine function (which is essentially the modular quantum dilogarithm) leads to a gauge/string duality hypothesis for an exceptional gauge groups. Particularly, this leads to values of GV integers for manifold after geometric transition. For complete proof of hypothesis one have to identify the manifold,  understand the interpretation of further terms  and appearance of even multicoverings, as well as present the string interpretation of other gauge-invariant quantities.

It is interesting to consider  an invariant volume of groups from \cite{M13} (which is essentially $Z_1$ with appropriate choice of $\delta$) on exceptional line. This can be interpreted as a generalization of matrix models \cite{Mar2} to exceptional groups. As shown by Ooguri and Vafa \cite{OV}, expansion of invariant volume on classical lines over $1/N$ has  coefficients which are virtual Euler characteristics of moduli space of surfaces of genus g with few crosscaps. In exceptional case one may show (in preparation) that similar expansion of volume on exceptional line has as $n$-th coefficients the sum of  $\zeta(n)/n$ and Bernoulli numbers with integer and rational coefficients, respectively (here $\zeta$ is Riemann's zeta function). One should further connect these values to characteristics of moduli space of Riemann surfaces.

\section{Acknowledgments.}

Work is partially supported by Volkswagen Foundation and  the Science Committee of the Ministry of Science and Education of the Republic of Armenia under contract  13-1C232. I'm thankful to S.Theisen and R. Minasian for discussions of present work and to H.Khudaverdian, H.Mkrtchyan and N.Reshetikhin for support at different stages of work. I'm indebted to Albert Einstein Institute (AEI MPI), where this work is finished, for hospitality,  to organizers of workshops "Low-dimensional topology and number theory" at MFO, Oberwolfach, and "Frontiers in field and string theory", Yerevan, for invitations and to participants for discussions.


\begin{thebibliography}{99}
\bibitem{M13}
R.L.Mkrtchyan, {\it Nonperturbative universal Chern-Simons theory},  JHEP09(2013)054, arXiv:1302.1507.

\bibitem{V0}
P.Vogel {\it Algebraic structures on modules of diagrams}, preprint (1995), 
J. Pure Appl. Algebra {\bf 215} (2011), no. 6, 1292-1339. 

\bibitem{H1}
G. 't Hooft, {\it A planar diagram theory for strong interactions}, Nucl.Phys. {\bf B72} (1974), 461-473.

\bibitem{W1} 
 E. Witten, {\it Quantum field theory and the Jones polynomial}, Comm. Math. Phys. {\bf 121} (1989),  351-399.
 
\bibitem{MV1}
R.L. Mkrtchyan and A.P. Veselov, {\it Universality in Chern-Simons theory}, JHEP08 (2012) 153, arXiv:1203.0766.

\bibitem{GV} 
R. Gopakumar and C. Vafa, {\it M-theory and topological strings, II}, arXiv:hep-th/9812127. 

\bibitem{M13-2} 
R.L.Mkrtchyan, {\it Universal Chern-Simons partition functions as quadruple Barnes' gamma-functions},  JHEP10(2013)190, arXiv:1309.2450

\bibitem{Nar}
Atsushi Narukawa, {\it The modular properties and the integral representations of the multiple elliptic gamma functions}, Adv. in Math. 189 (2) (2004) 247-267, arXiv:math/0306164 [math.QA]

\bibitem{Kur}
 N. Kurokawa, {\it Multiple sine functions and Selberg zeta functions}, Proc. Japan Acad. A 67
(1991) 61–64.

\bibitem{KurKoy}
N. Kurokawa, S.Koyama, {\it Multiple sine functions}, Forum Math. 15, (2003), 839-876.

\bibitem{Shi}
T. Shintani,{\it On a Kronecker limit formula for real quadratic fields}, J. Fac. Sci. Univ. Tokyo
24 (1977) 167-199.

\bibitem{SV}
S. Sinha and C. Vafa, {\it SO and Sp Chern-Simons at Large N}, arXiv:hep-th/0012136 (2000).

\bibitem{BFM}
Vincent Bouchard, Bogdan Florea and Marcos Marino, {\it Counting Higher Genus Curves with Crosscaps in Calabi-Yau Orientifolds}, JHEP0412:035,2004, arXiv:hep-th/0405083.

\bibitem{BFM2}
Vincent Bouchard, Bogdan Florea and Marcos Marino, {\it Topological Open String Amplitudes on Orientifolds}, JHEP0502:002,2005, arXiv:hep-th/0411227.

\bibitem{Del}
P. Deligne, {\it La s\'erie exceptionnelle des groupes de Lie}, C. R. Acad. Sci. Paris, S\'erie I {\bf 322} (1996), 321-326.

\bibitem {DM}
P. Deligne and R. de Man, {\it La s\'erie exceptionnelle des groupes de Lie II}, C. R. Acad. Sci. Paris, S\'erie I  {\bf 323} (1996), 577-582.  

\bibitem{LM1} 
 J.M. Landsberg and L. Manivel, {\it A universal dimension formula for complex simple Lie algebras}, Adv. Math. {\bf 201} (2006), 379-407.

\bibitem{Kne}
Jan Kneissler, {\it On spaces of connected graphs II: Relations in the algebra Lambda}, 
 Jour. of Knot Theory and its Ramif. vol. 10, no. 5 (2001), 667-674,  arXiv:math/0301019
 
\bibitem{Cvitbook}
P. Cvitanovic {\it Group Theory}, Princeton University Press, Princeton, NJ, 2008, 
http://www.nbi.dk/group theory

\bibitem{Cvi}
Predrag Cvitanovic,  {\it Negative dimensions and E7 symmetry}, Nucl. Phys. B188, 373 (1981).

\bibitem{CDM}
S. Chmutov, S. Duzhin and J. Mostovoy, {\it Introduction to Vassiliev Knot Invariants}, 
Cambridge University Press, (2012).

\bibitem{Pat}
B. Patureau-Mirand, {\it Caracteres sur l'algebre de diagrammes trivalents Λ}, Geom. Topol. 6 (20) (2002) 565-607.

\bibitem{D}
P. Deligne, (2013), unpublished.

\bibitem{MSV}
R.L. Mkrtchyan, A.N. Sergeev and A.P. Veselov,  {\it Casimir values for universal Lie algebra},  Journ. Math.Phys. 53, 102106 (2012),   arXiv:1105.0115.

\bibitem{Mkr2}
R.L.Mkrtchyan, {\it On a map of Vogel`s plane}, arxiv:1209.5709.

\bibitem{Barnes3}
E.W. Barnes, {\it On the theory of the multiple gamma function}, Trans. Cambridge Philos. Soc.
19 (1904), 374-425.

\bibitem{Rui}
S. N. M. Ruijsenaars, {\it On Barnes' Multiple Zeta and Gamma Functions}, Advances in Mathematics 156, 107-132 (2000).

\bibitem{Fad}
L. D. Faddeev, {\it Discrete Heisenberg-Weyl Group and Modular Group},
Lett.Math.Phys. 34 (1995) 249-254, arXiv:hep-th/9504111.

\bibitem{Fad2}
L. D. Faddeev, {\it Volkov's Pentagon for the Modular Quantum Dilogarithm},
Functional Analysis and Applications, vol.45(4), 2011, p.65, arXiv:1201.6464 [math.QA]. 

\bibitem{PS}
Sara Pasquetti and Ricardo Schiappa, {\it Borel  and Stokes Nonperturbative Phenomena in Topological String Theory and c=1 Matrix Models}, arXiv:0907.4082.


\bibitem{Mar1} 
Marcos Mari\~no, {\it Chern-Simons Theory and Topological Strings}, Rev.Mod.Phys.77:675-720, 2005, 
arXiv:hep-th/0406005.


\bibitem{BCOV}
M. Bershadsky, S. Cecotti. H. Ooguri, and C. Vafa,  { \it Kodaira-Spencer theory of gravity and exact results for
quantum string amplitudes}, Commun. Math. Phys. (1994) 165, 311, hep-th/9309140.

\bibitem{GP}
E. Getzler and R. Pandharipande,  {\it Virasoro constraints and the Chern classes of the Hodge bundle}, Nucl. Phys. (1998), B 530, 701,  math.AG/9805114.

\bibitem{FP}
C.Faber and R. Pandharipande,  { \it Hodge integrals and Gromov-Witten theory}, Invent. Math. (2000),139, 173, 
math.AG/9810173.

\bibitem{Vol}
 A. Yu. Volkov, {\it Noncommutative Hypergeometry}, Commun. Math.
Phys. 258 (2005) 257, arXiv:math/0312084 [math.QA].

\bibitem{Kash}
 R. M. Kashaev, {\it Quantization of Teichmueller spaces and the quantum
dilogarithm}, Lett. Math. Phys. 43, 105 (1998).

\bibitem{Mar2} 
Marcos Mari\~no, {\it Chern-Simons theory, matrix models and topological strings}, Clarendon Press, Oxford, 2005. 

\bibitem{OV}
H. Ooguri and C. Vafa, {\it Worldsheet derivation of a large N duality}, Nucl. Phys. {\bf B 641} (2003), hep-th/0205297.





\end{thebibliography}
\end{document}